\documentclass[12pt,preprint]{emulateapj}

\begin{document}

\newcommand{\chisq}{\chi^2}
\newcommand{\chisqdof}{\chi^2_\nu}
\newcommand{\re}{\tilde{r}_\oplus}
\newcommand{\tE}{t_{\rm E}}
\newcommand{\rEp}{r_{\rm E,p}}
\newcommand{\rE}{r_{\rm E}}
\newcommand{\thetaE}{\theta_{\rm E}}
\newcommand{\thetastar}{\theta_\star}
\newcommand{\tv}{\tilde{v}}
\newcommand{\tp}{t_{\rm p}}
\newcommand{\tc}{t_{\rm c}}
\newcommand{\RVI}{R_{\rm VI}}
\newcommand{\EVI}{E_{\rm V-I}}
\newcommand{\AV}{A_{\rm V}}
\newcommand{\VO}{V_{\rm 0}}
\newcommand{\MV}{M_{\rm V}}
\newcommand{\MI}{M_{\rm I}}
\newcommand{\VM}{V_{\rm M}}
\newcommand{\RM}{R_{\rm M}}
\newcommand{\ustar}{u_\star}
\newcommand{\phig}{\phi_\gamma}
\newcommand{\eps}{\epsilon}
\newcommand{\rad}{\thinspace{\rm rad}}
\newcommand{\nhat}{\hat{n}}
\newcommand{\xl}{x_{\rm L}}
\newcommand{\yl}{y_{\rm L}}
\newcommand{\zl}{z_{\rm L}}
\newcommand{\xlens}{x_{\rm L}}
\newcommand{\ylens}{y_{\rm L}}
\newcommand{\zlens}{z_{\rm L}}
\newcommand{\xearth}{x_\oplus}
\newcommand{\yearth}{y_\oplus}
\newcommand{\xearthl}{x_{\oplus, {\rm l}}}
\newcommand{\yearthl}{y_{\oplus, {\rm l}}}
\newcommand{\zearth}{z_\oplus}
\newcommand{\Drel}{D_{\rm rel}}
\newcommand{\thetarel}{\theta_{\rm rel}}
\newcommand{\murel}{\mu_{\rm rel}}
\newcommand{\pirel}{\pi_{\rm rel}}
\newcommand{\deltar}{\delta{\vec{r}}}
\newcommand{\Is}{I_{\rm s}}
\newcommand{\kpc}{{\rm\,kpc}}
\newcommand{\beqarr}{\begin{eqnarray}}
\newcommand{\eeqarr}{\end{eqnarray}}
\newcommand{\umin}{u_{\rm min}}
\newcommand{\tpeak}{t_{\rm peak}}
\newcommand{\Ibase}{I_{\rm base}}
\newcommand{\sigmal}{\sigma_{\rm l}}
\newcommand{\sigmab}{\sigma_{\rm b}}
\newcommand{\mul}{\mu_{\rm l}}
\newcommand{\mub}{\mu_{\rm b}}
\newcommand{\Dl}{D_{\rm l}}
\newcommand{\Ds}{D_{\rm s}}
\newcommand{\piE}{\pi_{\rm E}}
\newcommand{\pivec}{\mbox{\boldmath $\pi$}}
\newcommand{\piEl}{\pi_{\rm E, \parallel}}
\newcommand{\piEt}{\pi_{\rm E, \perp}}
\newcommand{\kms}{{\,\rm km\,s^{-1}}}
\newcommand{\mas}{{\rm \,mas}}
\newcommand{\masyr}{{\rm\mas}\,{\rm yr}^{-1}}
\newcommand{\rot}{{\rm rot}}
\newcommand{\Msol}{M_\odot}
\newcommand{\muvec}{{\boldsymbol\mu}}
\newcommand{\rvec}{\mathbf{r}}
\newcommand{\thetazero}{{\boldsymbol\theta_0}}
\newcommand{\uzero}{\mathbf{u}_0}

\shorttitle{First direct $\mu$-lens toward the bulge}
\shortauthors{}

\title{The first direct detection of a gravitational $\mu$-lens toward the Galactic Bulge\footnotemark[1]}

\footnotetext[1]{Based on observations made with the NASA/ESA {\it Hubble Space Telescope},
obtained from the data archive at the Space Telescope Science Institute.
STScI is operated by the Association of Universities for Research in Astronomy,
Inc. under NASA contract NAS 5-26555.
}

\author{
S.~Koz\l owski\altaffilmark{2},
P.R.~Wo\'zniak\altaffilmark{3},
S. Mao\altaffilmark{2} \&
A. Wood\altaffilmark{2}
}

\altaffiltext{2}{Jodrell Bank Observatory, University of Manchester, Macclesfield, Cheshire SK11 9DL, UK,
e-mail: simkoz@mlens.net}
\altaffiltext{3}{Los Alamos National Laboratory, MS-D466, Los Alamos, NM 87545}


\begin{abstract}
We present a direct detection of the gravitational lens that caused the microlensing event MACHO-95-BLG-37.
This is the first fully resolved microlensing system involving a source in the Galactic bulge, and the second
such system in general. The lens and source are clearly resolved in images taken with the High Resolution Channel
of the Advanced Camera for Surveys on board the {\it Hubble Space Telescope (HST)} $\sim9$ years after
the microlensing event. The presently available data are not sufficient for the final, unambiguous
identification of the gravitational lens and the microlensed source. While the light curve models combined with
the high resolution photometry for individual objects indicate that the source is red and the lens is blue,
the color-magnitude diagram for the line of sight and the observed proper motions strongly support
the opposite case. The first scenario points to a metal-poor lens with mass $M\approx0.6\Msol$
at the distance $\Dl\approx4$ kpc. In the second scenario the lens could be a main-sequence star
with $M=0.8$--$0.9\Msol$ about half-way to the Galactic bulge or in the foreground disk, depending
on the extinction.
\end{abstract}


\section{Introduction}
\label{sec:intro}

Gravitational microlensing of stars within the Local Group of galaxies (\citealt{BP96})
directly probes both luminous and dark matter concentrations along the line of sight.
Over the past decade microlensing surveys have continued to enable observations with
far-reaching implications, such as constraints on the fraction and content of Galactic
dark matter (e.g. \citealt{Alc96, Alc98, Alc01b}), discovery and characterization of exo-planet
systems (\citealt{Bon04,Uda05, Bea06, Gou06}), and measurements of the fundamental properties
of stars and their evolutionary end points (\citealt{Ben02, Abe03, Gou04}). Unfortunately,
while the light curve of a microlensing event provides the key discovery signature, it is
insufficient to solve uniquely for the mass, the distance and the relative transverse velocity of the lens.
As a result, out of a few thousand events discovered to date, only a handful allowed the mass
of the lens to be measured (\citealt{An02, Gou04, Jia04}).

In the case of microlensing by a luminous body (a star) the basic degeneracy of the model can be broken
by directly observing both the lens and the source. The difficulty with this approach, however,
is inherent in the geometry of microlensing that implies milli-arcsecond separations between
the lens and source components during the event. So far MACHO-LMC-5 was the only microlensing
event for which the lensing body has been resolved (\citealt{Alc01a}). The lens that gravitationally
magnified the source in the Large Magellanic Cloud turned out to be a nearby M dwarf in the Galactic
disk (\citealt{Dra04, Gou04}). \cite{Ben06} demonstrated the presence of a bright lens component
in the planetary microlensing event OGLE-2003-BLG-235/MOA-2003-BLG-53 and estimated the mass of the
host star using the centroid shift of the combined light.

Here we report a direct detection and mass measurement of the gravitational lens responsible for
the MACHO-95-BUL-37 event---the first fully resolved microlensing system involving a Galactic bulge
source, and the second such system in general.

\section{Microlensing event MACHO-95-BLG-37}
\label{sec:data}

The event was discovered by the MACHO collaboration as a single and apparently achromatic brightening
of object 109.20635.2193 in their photometric monitoring database of the Galactic bulge (see \citealt{Tho05}).
The object is quite faint ($V\sim20$ mag) and located in one of the densest fields covered by the survey:
equatorial (J2000) and Galactic coordinates $(\alpha,~\delta)$=(18h04m34.44s, $-$28$^\circ25'33.7''$),\break
($l,~b$)=(2$^\circ$\hskip-2pt.54, $-$3$^\circ$\hskip-2pt.33). The location of the MACHO-95-BLG-37 event
was one of the targets in our proper motion mini-survey of the Galactic bulge (\citealt{Koz06}). Each of the 35 fields
in the mini-survey was centered on a microlensed source and was covered by two {\it Hubble Space Telescope (HST)}
pointings taken several years apart. Using several relatively isolated stars we could co-register the {\it HST}
and ground-based MACHO images to within $0.1\arcsec$ and unambiguously identify microlensed sources,
even in the presence of additional stars that were only resolved in the {\it HST} images.

In the case of MACHO-95-BLG-37 we found that not only is the microlensed source accompanied by another very close star
with comparable brightness, but also that the relative proper motion of the two components places them within
$2.6\pm3.5$ mas of each other on 21 September 1995 (HJD 2449982.3) when the microlensing event took place. The prior
probability that the blend is a random coincidence is very small, so we have a clear indication
that the companion source is actually the gravitational lens that caused the event of 1995. Before we begin
a detailed investigation of this finding (\S\S\,\ref{sec:mlens_models}--\ref{sec:scenarios}) we first describe
the available data and basic data reductions.

\subsection{{\it HST} astrometry and photometry}
\label{sec:hst}

\begin{figure}
\includegraphics[width=\columnwidth]{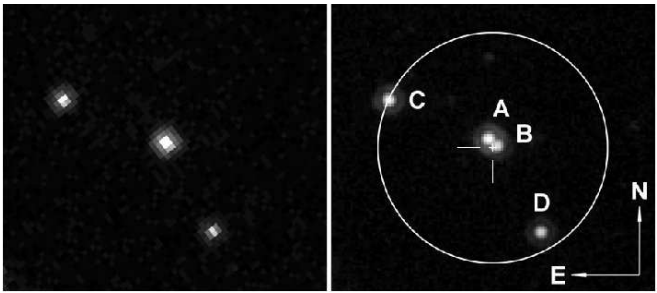}
\caption{{\it HST} images of the MACHO-95-BLG-37 event location.
The first epoch WFPC2/PC image (left) and the second epoch ACS/HRC image (right) were taken, respectively,
3.71 and 8.95 yr after the maximum light. The relative trajectory of stars A and B places them
within 2.6 mas of each other at the time of the microlensing event. The circle shows
a seeing disk characteristic of the ground-based MACHO survey (FWHM$\simeq2''\hskip-2pt .1$). 
The crosshair indicates the unbiased centroid of the lensed light from difference imaging.
\label{fig:fchart}}
\end{figure}

A detailed description of the relevant {\it HST} data\footnote{\tt http://archive.stsci.edu/hst/} was published
in \cite{Koz06} and only the essential facts are repeated here. The first and second epoch images were collected,
respectively, 3.71 and 8.95 yr after the event. The first pointing employed the Planetary Chip (PC)
of the Wide Field Planetary Camera 2 (WFPC2) instrument, and provided (nearly) simultaneous color information
in both $V$ and $I$ photometric bands using F555W and F814W filters. During the second pointing we used
the High Resolution Channel (HRC) of the Advanced Camera for Surveys (ACS) and obtained high signal-to-noise 
ratio (S/N) imaging
in the F814W filter only. In each case we co-added all suitable F555W and F814W images for a given epoch.
The field of view covered by the ACS/HRC and WFPC2/PC detectors is similar ($29\arcsec\times26\arcsec$
and $35\arcsec\times35\arcsec$, respectively), but not the pixel size ($25$ versus $45.5$ mas). At the {\it HST}
resolution the MACHO database object associated with the microlensing source was immediately revealed
to be a composite of four unresolved stars, which we label A through D (Fig.~\ref{fig:fchart}).

The magnitudes and positions of stars A--D were extracted from the fits of stellar profiles. The local point
spread function (PSF) models were generated using the {\small TINYTIM} software (\citealt{Kri93, Kri95})
and interpolated with bi-cubic splines. For all model fitting we used the {\small MINUIT}
package\footnote{\tt http://wwwasd.web.cern.ch/wwwasd/cernlib/}. Stars A and B have overlapping profiles
and required a special model with two PSF components fitted simultaneously. A small section of the ACS/HRC image
was fitted first, providing an unbiased value of the second epoch separation between the two components and a good
handle on the flux ratio in the $I_{\rm F814W}$-band. The flux ratio was then fixed at the second epoch value
for the purpose of fitting the $I_{\rm F814W}$-band WFPC2/PC image and obtaining the first epoch astrometry.
Finally, the $(V-I)$ colors of stars A and B were established by fitting the $V_{\rm F555W}$-band image using
a model with variable flux ratio and the blend separation fixed at the value taken from the $I_{\rm F814W}$-band
fit for the same epoch. The resulting astrometric and photometric measurements are given in Tables~\ref{tab:astrometry}
and \ref{tab:photometry}. Note that in the $V$-band the only available high resolution imaging comes from
the relatively shallow first epoch WFPC2/PC observation, so the A/B flux ratio is poorly constrained and the errors
in $V$ and $(V-I)$ are relatively large for these two stars. In the $I$-band, however, we have an accurate
measurement of the flux ratio from ACS/HRC
that allowed us to eliminate a degenerate free parameter from the double PSF fit of the WFPC2/PC data. This explains
why the astrometric accuracy for the first epoch is actually better than for the second epoch, despite a larger
pixel size and a much smaller separation between stars A and B in the WFPC2/PC images compared to the ACS/HRC data.
Using simulated images we found that our first epoch astrometry is biased by about 1.5\% toward lower separations.
We could not find a better procedure that would eliminate this effect, so a post-factum correction was included
in the WFPC2/PC data reported in Table~\ref{tab:astrometry}.

\vspace{0.3cm}
\subsection{Microlensing light curve revisited}
\label{sec:lc}

\begin{figure}
\includegraphics[width=\columnwidth]{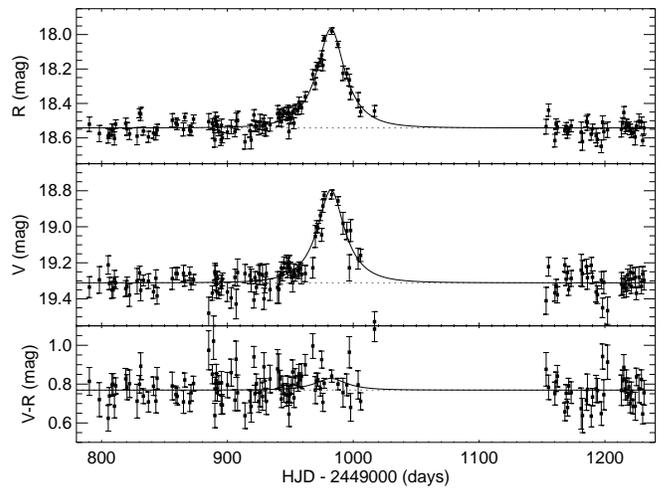}
\caption{Light curve of the MACHO-95-BLG-37 event from our Difference Image Analysis
of data in the public MACHO database. A significant color change near the peak magnification
indicates that the microlensed source is redder than the unresolved composite of stars A, B, C and D
from Fig.~\ref{fig:fchart}.
\label{fig:lc}}
\end{figure}

The MACHO-95-BLG-37 event was recorded in a faint star subject to intense crowding, and therefore 
the standard light curve in the MACHO photometric database has a 
very low S/N. In order to reduce the uncertainties of the microlensing parameters derived from light curve
modeling we performed Difference Image Analysis (DIA; \citealt{Ala98, Ala00, Woz00}) on the original 
ground-based images, i.e. on the simultaneous two-color imaging data collected by the MACHO
survey\footnote{\tt http://wwwmacho.mcmaster.ca/Data/MachoData.html}.
The PSF matching and photometric solutions were confined to a $2'\hskip-2pt .5\times2'\hskip-2pt .5$
region around the source (approximately $256\times256$ pixels). After discarding observations outside the relevant
time interval and rejecting a small fraction of frames with bad seeing we considered a total of 132 images
in each of the MACHO photometric bands $b_{\rm M}$ and $r_{\rm M}$. High S/N reference images were constructed
by co-adding 9 good quality images with a well-behaved PSF. From a series of difference frames in which the
source was significantly magnified we derived an unbiased centroid of the lensed light that clearly
points to the pair of stars A and B when transformed to the ACS/HRC coordinates (Fig.~\ref{fig:fchart}).
One of these two stars must then be the microlensed source.

\begin{deluxetable}{lccccc}
\tablewidth{\columnwidth}
\tablecaption{\label{tab:astrometry}{{\it HST} astrometry}}
\tablehead{
\colhead{Instrument} &
\colhead{} &
\colhead{Epoch} &
\colhead{$t-t_0$} &
\colhead{$\Delta\alpha$} &
\colhead{$\Delta\delta$}\\
\colhead{} &
\colhead{} &
\colhead{(yr)} &
\colhead{(yr)} &
\colhead{(mas)} &
\colhead{(mas)}
}
\startdata
WFPC2/PC  & & 1999.43 & 3.71 & $-29.9\pm1.3$ & $-24.9\pm1.3$ \\ 
  ACS/HRC & & 2004.67 & 8.95 & $-74.5\pm1.6$ &~$-63.1\pm1.6$
\enddata
\tablecomments{$\Delta\alpha$, $\Delta\delta$ are positions of star B relative
to star A (Fig.~\ref{fig:fchart}) in a Cartesian reference frame aligned
with the local equatorial coordinates. The moment of maximum light $t_0$
corresponds to HJD = 2449982.3 (Epoch 1995.72).
}
\end{deluxetable}

\begin{deluxetable}{cccccc}
\tablewidth{\columnwidth}
\tablecaption{\label{tab:photometry}{{\it HST} photometry}}
\tablehead{
\colhead{Star} &
\colhead{$V_{\rm F555W}$} &
\colhead{$I_{\rm F814W}$} &
\colhead{$(V-I)$} &
\colhead{$f_V$} &
\colhead{$f_I$}\\
\colhead{} &
\colhead{(mag)} &
\colhead{(mag)} &
\colhead{(mag)} &
\colhead{} &
\colhead{}
}
\startdata
A & $20.24\pm0.08$ & $18.45\pm0.02$ & 1.79 & 0.28 & 0.35 \\
B & $20.30\pm0.08$ & $19.07\pm0.04$ & 1.23 & 0.26 & 0.19 \\
C & $20.29\pm0.03$ & $18.66\pm0.03$ & 1.63 & 0.27 & 0.28 \\
D & $20.63\pm0.04$ & $19.18\pm0.04$ & 1.45 & 0.19 &~0.18
\enddata
\tablecomments{$f_V, f_I$ are fractional contributions to the total flux}
\end{deluxetable}

The reference flux in each band was derived from a comparison between our differential fluxes
and conventional PSF photometry obtained with the Dophot software (\citealt{Sch93}) running in a 
fixed-position mode with the input object lists based on our deep reference images. Stars C and D could not be
properly deblended, even using fixed {\it HST} positions transformed to the template coordinates.
The template position of the A--D composite was set to the mean ACS/HRC position of stars A and B.
We selected 31 calibration images per photometric band with the best overall seeing, background and
transparency. In seven of these images the source was visibly magnified. The statistical uncertainty
of the reference flux is 8\% in $b_{\rm M}$ and 9\% in $r_{\rm M}$. The background level estimated
by the Dophot algorithm in a crowded field is somewhat sensitive to the assumed shape of the PSF
(especially in the wings). In our case of a very faint object near the detection limit set by the source
confusion we find that the systematic uncertainty in the reference flux can easily reach 10\%.
This generic problem is partially alleviated by the fact that the systematics are similar in both
filters and source blending must always be considered in the analysis of individual light curves
in crowded fields.

The final light curves (Fig.~\ref{fig:lc}) were shifted to the instrumental $b_{\rm M}$, $r_{\rm M}$ scale
of the MACHO database using a median offset for a few tens of bright stars near the location of the MACHO-95-BLG-37
event and transformed to approximately standard $\VM, \RM$ magnitudes following \cite{Pop05}.
We also determined transformations between $\VM, \RM$ and the standard $V, I$ magnitudes:

\begin{equation}
\begin{array}{ccr}
\VM&=& V + (0.05\pm0.11)(V-I) + (0.01\pm0.20), \\
\RM&=& I + (0.62\pm0.10)(V-I) - (0.17\pm0.18).
\end{array}
\label{eq:filters}
\end{equation}

\noindent
Hereafter, the subscript is omitted and MACHO filters are implied for $V, R$ photometry.
The overall quality
of the $V$-band light curve is lower compared to that in the $R$-band due to occasional pixel level defects in the
$b_{\rm M}$ frames that were clearly visible in the difference images.

\section{Microlensing light curve models}
\label{sec:mlens_models}

The first step is to obtain the basic microlensing parameters such as the time-scale $\tE$, the dimensionless impact
parameter $u_0$, the moment of the peak brightness $t_0$, and the baseline magnitudes $m_{V, R}$. In order to preserve
consistent color information, both $V$- and $R$-band light curves were fitted simultaneously with a simple microlensing
model that allows for flux blending (source fractions $f_{\rm s} < 1$). The data point at $t=1016.9$ days is a moderate
outlier in the $V$-band light curve (Fig.~\ref{fig:lc}) and is rejected in all analyses. The change in $\chisq$ due to
this cosmetic change is not significant and none of our conclusions are affected. The resulting best fit model is given
in Table~\ref{tab:lc_fits} and provides a marginally acceptable fit (reduced $\chisq_\nu=1.49$ for $\nu=255$ degrees of freedom).

\begin{deluxetable}{lcccccc}
\tablewidth{8cm}
\tablecaption{\label{tab:lc_fits}{Microlensing light curve model}}
\tablehead{
\colhead{\makebox[2.5cm][l]{Parameter}} &
\colhead{\makebox[2.0cm][c]{Value}} &
\colhead{\makebox[2.0cm][c]{Error}}
}
\startdata
$t_0$ (days)          \dotfill & 982.3   & 0.3     \\
$\tE$ (days)          \dotfill & 25.2    & 4.2     \\
$u_0$                 \dotfill & 0.37    & 0.10    \\
$f_{{\rm s},V}$       \dotfill & 0.33    & 0.12    \\
$f_{{\rm s},R}$       \dotfill & 0.38    & 0.14    \\
$m_V$ (mag)           \dotfill & 19.314  & 0.005   \\
$m_R$ (mag)           \dotfill & 18.545  & 0.003   \\
$\chisqdof$           \dotfill & 1.490   & \nodata \\
$\nu$                 \dotfill & 255     & \nodata \\
\enddata
\tablecomments{Maximum magnification is at $t_0$ days after HJD = 2449000.}
\end{deluxetable}

\subsection{Colors}
\label{sec:colors}

Using different parameterizations of the model equivalent to the one in Table~\ref{tab:lc_fits}, we obtained
the source/blend colors and the color difference with the error bounds that fully account for covariance:
$(V-R)_{\rm s}=0.92\pm0.04$, $(V-R)_{\rm b}=0.68\pm0.04$ and
$\Delta (V-R)_{\rm s, b}=(V-R)_{\rm s}-(V-R)_{\rm b}=0.24\pm0.06$. This corresponds to a positive color shift
during the event $\Delta (V-R)_{\rm event}\simeq+0.06$ mag and indicates that the source is redder than the blend. However,
it must be emphasized that the measurement of the reference flux for our light curves poses a significant challenge
given the limitations of the available archival data (\S\,\ref{sec:lc}). Both $(V-R)_{\rm b}$ and
$\Delta (V-R)_{\rm s, b}$ are subject to the systematics of the reference flux in two bands. The value of
$(V-R)_{\rm s}$, on the other hand, is more reliable, because it is constrained by the magnified portion
of the light curve, even if the reference fluxes are not known.
This is best seen from the model of the simultaneous two-color DIA light curve written as
$\Delta F(t) = F_{\rm s}\times A(t) + F_0$ in each band,
where $A(t)$ is the magnification factor and $F_0 < 0$ if the source is effectively magnified in the reference image.
Although in most cases the source flux $F_{\rm s}$ is poorly constrained in both colors,
the error bounds on the ratio $F_{{\rm s}, V}/F_{{\rm s}, R}$ are relatively tight due to covariance and,
most importantly, independent of the flux offsets. Therefore, the derived value of $(V-R)_{\rm s}$ only depends
on the global calibration of flux units for the reference images, which can be done much more reliably using bright
isolated stars.
In conjunction with the {\it HST} photometry, the source color information will be crucial to deciding the identity
of the microlensed source, and therefore the lens (\S\,\ref{sec:lens}).

\subsection{Parallax constraints}
\label{sec:parallax}

The ground-based microlensing light curve provides useful constraints on the acceleration term in the
observed trajectory of the lens relative to the source.
In the case of a short, low-magnification microlensing event such as MACHO-95-BLG-37 we can only obtain one-dimensional
information (\citealt{Gou94}). Following the geocentric formalism of \cite{G04}, we introduce into the model
the dimensionless microlensing parallax vector $\pivec_{\rm E}$, where $\piEl$ is the component of $\pivec_{\rm E}$ opposite the
direction of the projected position of the Sun at the peak of the event. We find that $\piEl = 0.07^{+0.65}_{-0.46}$,
while $\piEt$ remains unconstrained, i.e. there is no detectable parallax. There are two observations during the event
(at $t=967.9$ and $t=996.9$ days) with atypically low $V$-band fluxes and relatively large error bars compared to the
adjacent measurements. Without these two data points we get $\piEl=0.00^{+0.67}_{-0.45}$ and the apparent weak asymmetry
of the best fit model goes away. In \S\,\ref{sec:lens} this constraint is improved using ${\it HST}$ photometry
and in \S\,\ref{sec:rE} combined with the ${\it HST}$ astrometry to place the limits on the relative source-lens
parallax $\pirel$.

\section{Resolution of the microlensing system into lens and source}
\label{sec:resolution}

The fundamental difficulty with resolving a lens detected through time-variable magnification is that its
apparent separation from the source is below the {\it HST} resolution for months or even years after the event.
In the case of MACHO-95-BLG-37 (and similarly for MACHO-LMC-5) this problem is greatly reduced due to the rather large
relative motion of stars A and B (\S\,\ref{sec:hst}). High precision {\it HST} astrometry at two epochs
well after the peak magnification allowed us to calculate a very accurate relative trajectory of star B
with respect to star A. Simply connecting the two measurements in Table~\ref{tab:astrometry} we get:

\begin{equation}
\rvec(t) = \rvec(t_1){(t_2-t)\over(t_2-t_1)} + \rvec(t_2){(t-t_1)\over(t_2-t_1)},
\label{eq:trajectory}
\end{equation}

\noindent
where $\rvec = (\Delta\alpha, \Delta\delta)$ is the relative position with measurements available at $t_1=3.71$ and $t_2=8.95$ yr.
The separation at the peak of the event ($t=0$) is then:

$(\Delta\alpha_0, \Delta\delta_0) = (1.6\pm2.5,~2.1\pm2.5)~{\rm mas}.$


\noindent
These values are fully consistent with a model in which the two stars are the source and lens, and which 
predicts a very low value of the two-dimensional separation $r_0 = r(0)$. There are two
alternative possibilities: either one of the members of the pair is a random interloper, or it is a companion to 
either the lens or the source. The first possibility is ruled out by the following argument: In the sky region under 
consideration the density of stars is 0.085 and 0.176 per square arcsecond for stars brighter than
$I=18.45$ and 19.07 mag, respectively. The corresponding Poisson probabilities of a random alignment
within 2.6 mas at the time of the event are $1.8\times10^{-6}$ and $3.7\times10^{-6}$, respectively, i.e. very low. 
The other case, of one of the two detected stars being a companion to
either the lens or the source, can also be ruled out.
It is clear is that one of the two stars must be the source. Furthermore,
the rapid relative proper motion excludes the possibility that the second star is a companion 
of the source (the implied binary motion will be too high, about 400 km/s at a distance of 8\,kpc). Thus we only 
need to consider the possibility that one of the stars is a companion of
an unseen dark lens, with a separation of about 2.6 mas between them
(recall that the lens is almost perfectly aligned with the lensed source at the peak).
This is about $b\approx 3.5$ Einstein radii (for $\thetaE=0.75\pm
0.13$ mas, see \S4.2). We can approximately model the perturbation of the
luminous star on the dark lens as a Chang-Refsdal lens (Chang \& Refsdal 1984).
The shear induced by the luminous star at the position of the dark lens
would be $\gamma=q/b^2 \approx 0.08$, where $q$ is the mass ratio
of the luminous companion to the dark lens. The short $\tE$ does not
favor a massive dark lens such as a black hole and neutron star, and so
the mass ratio $q$ is likely larger than one.
The caustics will have a size roughly $2\gamma/\sqrt{1-\gamma} \approx 2q/b^2 \sim
0.16$ (e.g. Mao 1992). The caustics size is comparable to the measured 
impact parameter ($u_0 \sim 0.37$), which would introduce a strong
asymmetry in the light curve for most trajectories (not seen 
in the observed low S/N light curves). Hence we regard the `dark' lens scenario
as not very likely. The bright lens hypothesis is thus favored and we conclude that the source and lens
system involved in the MACHO-95-BLG-37 microlensing event consists of stars A and B from Figure~\ref{fig:fchart}
(in an order still to be determined). Our subsequent arguments are based on that assumption.


\subsection{Identifying the lens}
\label{sec:lens}

\begin{figure}
\includegraphics[width=\columnwidth]{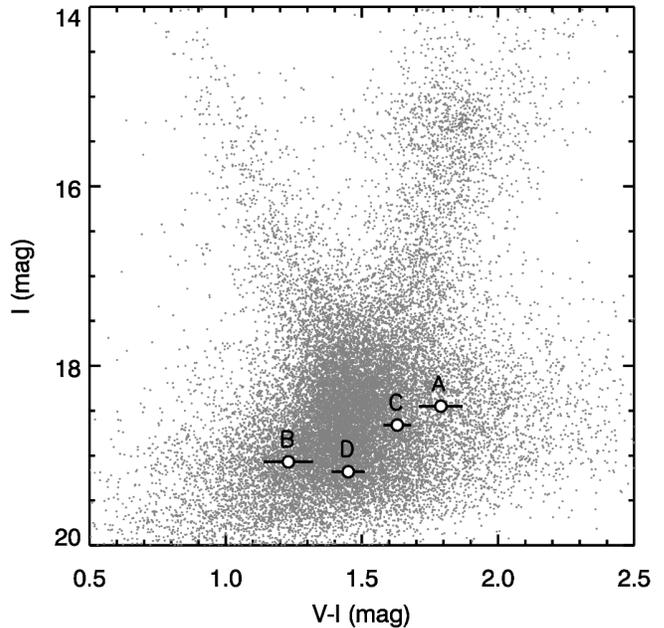}
\caption{Color-magnitude diagram (CMD) of stars around the MACHO-95-BLG-37 event location.
The {\it HST} photometry of stars A--D from Fig.~\ref{fig:fchart} (open circles) is plotted
against the average bulge and disk population along the line of sight (OGLE-II data; \citealt{Uda02}).
\label{fig:cmd}}
\end{figure}

To find out which member of the candidate pair of stars is the lens, we can make use of the observed color change
during the event and the fact that gravitational lensing is achromatic. In Figure~\ref{fig:cmd} we plot
the color-magnitude diagram (CMD) of the stellar field around MACHO-95-BLG-37 and stars A--D from
Figure~\ref{fig:fchart}. The light curve in Figure~\ref{fig:lc} reflects the integrated flux of the four stars
(unresolved in ground-based images). From the observed color increase $\Delta (V-R)_{\rm event}\simeq+0.06$ mag
near the peak magnification we infer that the microlensed source is redder than the composite. Although
the $(V-R)$ colors were not measured individually for stars A--D, it is very unlikely that the ordering
of the $(V-I)$ and $(V-R)$ colors is different. Stars A and B are, respectively, the reddest and the bluest
components of the blend, so the color shift points to star A as the source. This is entirely in agreement
with the source color $(V-R)_{\rm s}=0.92\pm0.04$ mag and the color difference $(V-R)_{\rm s, b}=0.24\pm0.06$ mag
between the source and the rest of the blend found in \S\,\ref{sec:colors}. Equation~(\ref{eq:filters}) implies 
{\it HST} $(V-I)_{\rm s}\simeq1.7$ mag, also consistent with the {\it HST} photometry of star A.
Thus, based on the light curve evidence, star B must be the lens, because star A is the microlensed source.
However, in \S\,\ref{sec:scenarios} we show that the physical interpretation of the CMD and kinematic data
strongly argues against this result.

In principle, the {\it HST} photometry provides an additional test
of these possibilities because we can transform the measurements to the MACHO system and obtain a constraint
on the source magnitudes $V=20.34\pm0.09, R=19.38\pm0.05$ assuming star A and $V=20.37\pm0.11, R=19.66\pm0.08$
assuming star B (including the variance and covariance in transformation coefficients).
Unfortunately, the difference $\Delta \chisq\simeq0.1$ between the models with stars A and B as the source 
is completely insignificant. An additional problem is that the contributions of stars A--D to the blend
are not known very well. In order to match the total MACHO baseline magnitudes (\S\,\ref{sec:lc}) we would have to
make the transformed {\it HST} fluxes of all four stars fainter by 10--15\%, depending on the photometric band,
and then still assume that only about half of the flux in stars C and D is effectively added to the total
flux of stars A and B.
This is not surprising knowing that stars C and D are near the edge of the FWHM disk of stars A and B
(Fig.~\ref{fig:fchart})---yet they are too faint to be deblended---and that in ground-based microlensing images
the ``sky'' level is set by a featureless continuum of merging stars. Overestimating the background by a mere
few counts makes stars near the detection limit appear 0.1--0.2 mag fainter. The weights are probably
slightly different in each photometric band due to details such as the orientation of stars C and D
with respect to the PSF that is never perfectly round. Nevertheless, it is still useful to perform
a microlensing light curve fit with a single additional ``measurement'' of the source magnitude,
i.e. effectively constrain $f_{\rm s}$. Compared to the results in \S\,\ref{sec:parallax}, the error bounds
on the dimensionless parallax are improved, yielding $\piEl=0.0\pm0.4$ for any reasonable set of flux weights.

\subsection{Proper motion, Einstein ring radius and relative parallax}
\label{sec:rE}

The relative proper motion $\murel$ of the lens with respect to the source provides further clues
about the physics of the event. For the case at hand, a robust value of $\murel$ can be derived
from the {\it HST} astrometry alone. If we ignore the parallax and approximate $u_0=0$, we find
$\murel=10.85\pm0.16\masyr$ and the position angle $\phi=229.83\pm0.84$ deg (North through East).
Allowing for a finite impact parameter makes no difference to $\murel$, changes $\phi$
by a mere $0.4$ deg, and predicts the peak brightness a couple of months after the actual event,
consistent within 1$\sigma$ uncertainties. Including the parallax also has a negligible
influence on the trajectory. Thus the Einstein radius can be estimated as
$\thetaE=\murel\tE=0.75\pm0.13$ mas.

The direction of the dimensionless parallax vector $\pivec_{\rm E}$ is the same as the direction of the
lens-source relative proper motion. The component $\piEl$ that points away from the projected position
of the Sun is almost perfectly due East, since the event peaked on 21 September. Using results from
\S\,\ref{sec:parallax} and \S\,\ref{sec:lens} we can immediately estimate $\piE = |\piEl(\sin\phi)^{-1}| < 0.53$,
and set an upper limit on the relative lens-source parallax, $\pirel=\piE \thetaE \la 0.3$ mas.

\section{Microlensing scenarios and the lens mass}
\label{sec:scenarios}

\begin{deluxetable}{lcc}
\tablewidth{\columnwidth}
\tablecaption{\label{tab:kinematics}{{\it HST} kinematics of stars A and B}}
\tablehead{
\colhead{star} &
\colhead{$\mul$} &
\colhead{$\mub$}\\
\colhead{} &
\colhead{($\masyr$)} &
\colhead{($\masyr$)}
}
\startdata
A & \makebox[1.3cm][r]{$6.9\pm 0.3$} & \makebox[1.3cm][r]{$-0.2\pm0.3$} \\ 
B & \makebox[1.3cm][r]{$-3.8\pm0.4$} & \makebox[1.3cm][r]{$3.7\pm0.4$} \\
\enddata
\tablecomments{Proper motions are expressed in an average star reference frame as defined in \cite{Koz06}.}
\end{deluxetable}

The value of the source color derived in \S\,\ref{sec:lens} favors a scenario in which star A is the source
and star B is the lens. However, as we show in this section, such arrangement is very unlikely in the context
of the CMD (Fig.~\ref{fig:cmd}) and proper motions measured relative to the Galactic bulge (Table~\ref{tab:kinematics}).
Although the extinction-to-reddening ratio in the direction of the event is abnormally low, the reddening anomaly
cannot explain the conflict. After \cite{Sum04}, we adopt the reddening coefficient $\RVI = \AV/\EVI = 1.98$
and the total bulge extinction $\AV=1.54$ mag. In the following discussion we consider both source star cases 
in some detail and then use the measurement of $\thetaE$ from \S\,\ref{sec:rE} to constrain the mass
of the lens.

\subsection{Blue lens scenario}
\label{sec:blue_lens}

First we attempt to reconcile all available data with the evidence in \S\,\ref{sec:lens} that the source is red.
Given its red color, star A is too faint to be a giant and too bright
to be on the main sequence in the Galactic bulge. If it were a giant several magnitudes behind the bulge,
it could in principle belong to the Sgr dwarf galaxy, but its observed proper motion is not consistent
with Sgr (\citealt{Iba97}). More likely, star A is a dwarf in the foreground disk at a distance of
$\sim2.5$ kpc and behind most of the extinction. Then, if star B is indeed the lens, it must be in front
of star A, and the only simple solution is that the lens is a nearby white dwarf at $\sim100$ pc or so.
Unfortunately, this exciting possibility is ruled out by the parallax constraint $\pirel < 0.3$ mas
(\S\,\ref{sec:rE}), as it predicts $\pirel\sim10$ mas.

The location of star A in the CMD is still marginally consistent with a faint subgiant on the far side
of the bulge subject to $\sim0.2$ mag of extra reddening compared to the general population. But there
is little support for that, since the CMD shows a compact red clump and indicates a very uniform
extinction across this field (c.f. the extinction map of \citealt{Sum04}). The observed kinematics 
would also be very unusual for this scenario with
star A showing a 7 $\masyr$ disk-like prograde motion in the plane and star B moving at a $\sim135^\circ$
inclination. One could still argue that star B is a low metallicity halo subdwarf to explain
its motion and dramatically increase the $\AV$ prediction, but there is simply too much fine
tuning to consider this a reliable solution.

\subsection{Red lens scenario}
\label{sec:red_lens}

The properties of both stars are much easier to explain if we dismiss for a moment the source color evidence
from \S\,\ref{sec:lens} and assume that star B is the source and star A is the lens. In this case star B
is most likely in the bulge, where its absolute magnitude and color would be approximately $\MI=3.6, (V-I)_0=0.4$
assuming a red clump at $I_{\rm RC}=15.3, (V-I)_{\rm RC}=1.8$ in Figure~\ref{fig:cmd} and adopting
$M_{\rm I, RC}=-0.2, (V-I)_{\rm 0, RC}=1.0$ (\citealt{Uda00}). So the source fits the properties
of a metal-poor star near the turnoff point in the bulge, and the observed proper motion is fully
consistent with this picture. Then star A must be the lens and can be placed on the main sequence
at a distance of $\sim4$ kpc, where it would follow the Galactic rotation near the plane and move
a few $\masyr$. Again, the observed kinematics support this scenario.

\subsection{The lens mass estimates}
\label{massestimates}

Any acceptable scenario for the lens must satisfy the constraint on microlensing geometry set by
the measurement of the Einstein radius $\thetaE=0.75\pm0.13$ mas (\S\,\ref{sec:rE}). For the lens
of mass $M$ we have

\begin{equation}
\label{eq:thetaE_constrain}
M = {\thetaE^2 \over \kappa\pirel} \Msol, ~~~\kappa\simeq8.14~{\rm mas},
\end{equation}

\noindent
where $\pirel=\Dl^{-1} - \Ds^{-1}$ is the relative parallax for the lens and source distances
$\Dl$, and $\Ds$ kpc. A given value of the source distance $\Ds$ sets a relationship between
the lens mass $M$ and the range of lens distances allowed by the error bounds of $\thetaE$.
Making a reasonable assumption about the luminosity class of the lens we can parameterize the photometric
solutions in the same way, i.e. using the mass of the lens. 

\begin{figure}
\includegraphics[width=\columnwidth]{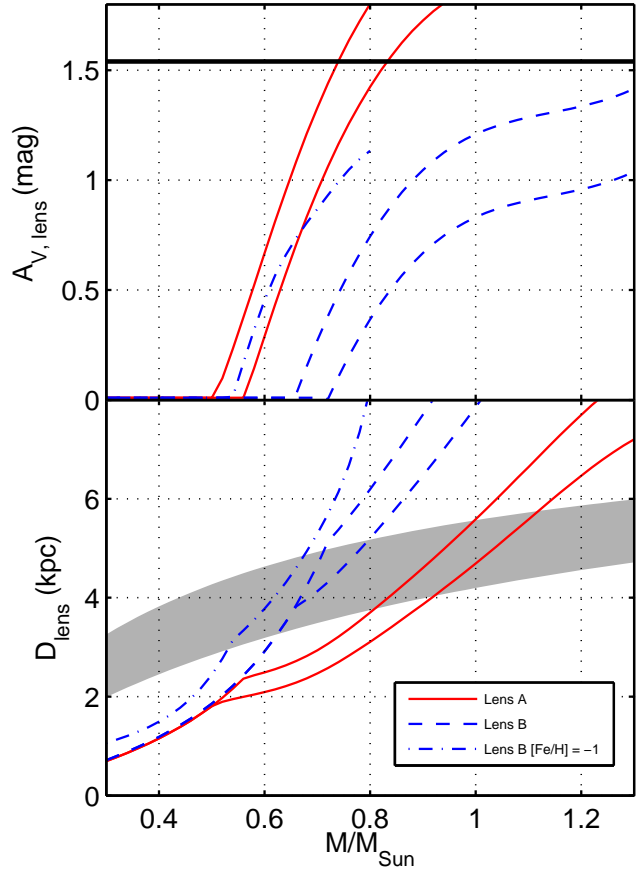}
\caption{Photometric solutions for the lens distance and
required extinction (top panel) as a function of the lens mass are shown.
The gray area covers microlensing geometries for a source at 8 kpc
allowed by the measurement of the Einstein radius $\thetaE=0.8$ mas
with 20\% uncertainty. For each main-sequence model we show a pair 
of lines that reflects the $\pm0.1$ mag uncertainty in stellar colors.
A metal-poor model is also shown for star B (dot-dashed line).
The top panel shows the required extinction to match the observed magnitude and
color of the lens (\S\ref{massestimates}); the thick horizontal line
marks the total Galactic extinction ($A_{\rm V}=1.54$ mag) for the line of sight from \cite{Sum04}.
\label{fig:solution}}
\end{figure}

For each value of the lens mass $M$ we use the appropriate
mass-luminosity-color relation to obtain the absolute magnitude $\MV$ and color $(V-I)_0$.
Then using the {\it HST} photometry (\S\,\ref{sec:hst}) we estimate the reddening $\EVI$ through
$\EVI=(V-I)_{\rm HST}-(V-I)_0$, and extinction $\AV$ using the reddening coefficient $\RVI = \AV/\EVI = 1.98$
taken from the extinction map of \cite{Sum04}. Combined with the assumption of the source located at 8 kpc,
each set of the above parameters allows a calculation of the lens distance $\Dl$
and extinction $\AV$, which should not exceed total extinction of the bulge $\AV=1.54$ mag
(see Fig.~\ref{fig:solution}).
We adopted the mass-luminosity relation for the main sequence from \cite{SchK82}
and the empirical color-magnitude relation defined by our polynomial fit to the {\it Hipparcos} CMD data in absolute
magnitudes ({\it Hipparcos} catalogue; \cite{Per97}, \cite{Bes90}, compiled by
I.\,N. Reid\footnote{\tt http://www-int.stsci.edu/$\sim$inr/cmd.html}).
Our CMD locus for the main sequence is very close to the linear relation of \cite{Rei91} for
$0.5 \Msol < M < 1.0 \Msol$ and is brighter by up to 0.3--0.5 mag outside this range. For comparison
we also used a model grid of low metallicity hydrogen-burning stars with [Fe/H] = $-$1.0 from \cite{Bar97}.

The constraints on the lens resulting from the two major scenarios are plotted in Figure~\ref{fig:solution}.
The red solid lines show the photometric constraints for star A being the lens and the 
blue dashed lines are for star B being the lens. Each pair of lines
corresponds to a range of solutions allowed by the $\pm0.1$ mag uncertainty in stellar colors. The gray
area is the geometric constraint based on the measurement of $\thetaE$ (eq.~[\ref{eq:thetaE_constrain}])
assuming a Galactic bulge source. It is clear that a blue lens (star B) on the main sequence generally underpredicts
the amount of extinction for a given distance. If the source is in the bulge, the microlensing constraint
selects $\Dl$=4--5 kpc, where the blue color of the lens does not allow any significant extinction. Moving
the source to a distance much larger than $\Ds=8$ kpc shifts the range of $\Dl$ upward by a few kpc,
but the problem with $\AV$ remains. The blue dot-dashed line illustrates the effect of lowering the metallicity
of star B to [Fe/H] = $-$1.0. This does solve the issue of extinction, but requires that the lens
is a metal-poor subdwarf in the Galactic halo, or perhaps in the thick disk. Such a possibility is unlikely
since the halo and thick disk contribute only a small fraction of stars within the Galactic disk.
The mass and the distance of the lens are then $M\sim0.6\Msol$ and $\Dl\sim4$ kpc.
The solution with a red lens (star A) is also not without a wrinkle, because in order to avoid
overshooting the total extinction for the bulge we need to make $\thetaE$ about 20\% larger
and the lens $\sim0.1$ mag bluer compared to the best estimates. Nevertheless,
we can still find a consistent answer within $1\sigma$ uncertainties. In this scenario the lens
is a main-sequence star with $M=0.8$--$0.9\Msol$ (spectral type G5--K0) at a distance of $\Dl\approx4$ kpc.

\section{Discussion}
\label{sec:discussion}

There is little doubt that we are directly observing the lens in the MACHO-95-BLG-37 event as it separates
from a nearly perfect alignment with the microlensed source. However, the final identification
of the gravitational lens is somewhat problematic. While the light curve models combined with
the photometric data for individual
objects favor a scenario with a blue lens and a red source, the opposite assignment is much more
plausible in the context of the color-magnitude diagram for the line of sight and the observed
proper motions. In any case,
the lens is a relatively bright star with the mass of $\sim0.6\Msol$ or $\sim0.9 \Msol$. It is conceivable that
additional factors such as binarity of stars affect the interpretation of the MACHO-95-BLG-37 event,
but a conclusive resolution of the present conflict will require new data. High quality spectroscopy
would unambiguously pinpoint both the 3-D kinematics and the distance scale.

For the first time in a Galactic bulge event the lens and source have been 
directly resolved. This is an important addition to the sample of one consisting of
the MACHO-LMC-5 event. There are thousands of known Galactic bulge microlensing events
and several dozen of those have archival {\it HST} pointings suitable for follow-up proper motion work.
In our proper motion mini-survey (\citealt{Koz06}) we included 35 of those fields, and have already
identified several more promising candidate lenses. 
As pointed out by \cite{HC03} and \cite{W06}, in a few per cent of microlensing events
toward the Galactic center a lens with characteristic motion $\murel\lesssim10\masyr$
may be detectable a decade after the microlensing episode.
We are only beginning to probe directly the mass spectrum of the Galactic microlenses;
however, we can expect that in the short term the progress will accelerate considerably
due to availability of the archival and future {\it HST} pointings.

\acknowledgments
We thank Prof. Andy Gould, the referee, for many useful comments that helped
us to improve this paper. We also thank Drs. Thomas Vestrand, {\L}ukasz Wyrzykowski and Martin Smith
for helpful discussions.

Support to P.R.W. for the proposal SNAP-10198 was provided by NASA through a grant from the STScI,
which is operated by the AURA under NASA contract NAS5-26555. This work was partly supported by
the European Community's Sixth Framework Marie Curie Research Training Network Programme, Contract
No. MRTN-CT-2004-505183 `ANGLES', in particular S.K. through a studentship and S.M. and A.W. through travel
support.

\end{document}